\documentstyle[amssymb,aps]{revtex}

\begin{document}
\title{Fourth SM Family Manifestations at CLIC}
\author{R. \c{C}ift\c{c}i}
\address{Physics Dept., Faculty of Sciences and Arts, Gazi University, 06500\\
Teknikokullar, Ankara, Turkey}
\author{A. K. \c{C}ift\c{c}i and E. Recepo\u{g}lu}
\address{Physics Dept., Faculty of Sciences, Ankara University, 06100 Tandogan,\\
Ankara, Turkey}
\author{S. Sultansoy}
\address{Physics Dept., Faculty of Sciences and Arts, Gazi University, 06500\\
Teknikokullar, Ankara, Turkey\\
Institute of Physics, Academy of Sciences, H. Cavid Ave. 33, Baku, Azerbaijan}
\maketitle

\begin{abstract}
The latest electroweak precision data allow the existence of additional
chiral generations in the standard model. We \ study prospects of search for
the fourth standard model family fermions and quarkonia at $e^{+}e^{-}$ and $%
\gamma \gamma $ options of CLIC. It is shown that CLIC will be powerfull
machine for discovery and investigation of both fourth family leptons and
quarkonia. Moreover, the formation of the fourth family quarkonia will give
a new opportunity to investigate Higgs boson properties.
\end{abstract}

\section{Introduction}

\noindent

Today, the mass and mixing patterns of the fundamental fermions are the most
mysterious aspects of particle physics. Even, the number of fermion
generations is not fixed by the Standard Model (SM). In this sense, SM may
be deliberated as an effective theory of fundamental interactions rather
than fundamental particles. The statement of the flavor democracy (or, in
other words, the Democratic Mass Matrix (DMM) approach), which is quite
natural in the SM framework, may be considered as the interesting step in
the true direction \cite{Harari,Fritzsch87,Fritzsch90,Fritzsch94}. It is
intriguing that, flavor democracy favors the existence of the fourth
standard model family \cite{Datta,Celikel,Atag,Sultansoy}. The lower limits
to masses of the fourth family fermions are \cite{Groom}: $m_{\nu
_{4}}\gtrsim 50$ GeV from LEP 1, $m_{l_{4}}\gtrsim 100$ GeV from LEP 2, $%
m_{d_{4}}>199$ GeV (neutral current decays, $d_{4}\rightarrow qZ$) and $%
m_{d_{4}}>128$ GeV (charged current decays, $d_{4}\rightarrow qW$) from FNAL
(Tevatron Run I).

Recently \cite{He} it is shown that a single extra chiral family with a
constrained spectrum is consistent with latest presicion data without
requiring any other new physics source. Then, as quoted from a recent paper 
\cite{Okun}: ''It is shown that additional chiral generations are not
excluded by the latest electroweak precision data if one assumes that there
is no mixings with the known three generations. In the case of ''heavy extra
generations'', when all four new particles are heavier than Z boson, quality
of the fit for the one new generation is as good as for zero new generations
(Standard Model)''. It should be noted that, a degenerate fourth SM family
is still disfavored by the data. Although the number of families for
degenerate case ($T=U=0$) is $2.97\pm 0.30$, this restriction can be relaxed
by allowing $T\neq 0$ (which means either non-degenerate extra generations
or a new physics), for example, $N_{fam}=3.27\pm 0.45$ for $T=0.10\pm 0.11$ 
\cite{Erler,Langacker}

The fourth family quarks will be copiously produced at the LHC \cite
{Arik,ATLAS} if their masses are less than $1$ TeV. Concerning the Tevatron
Run II, its higher luminosity will increase the discovery limits of $d_{4}$
quarks with respect to Tevatron Run I data approximately 50 \%. In addition,
the Higgs boson ''golden mode'' ($H\rightarrow ZZ\rightarrow 4l$, where $%
l=e,\mu $) will be observable at the upgraded Tevatron for $125<m_{H}<165$
GeV and $175<m_{H}<300$ GeV with more than $3$ $\sigma $ significance \cite
{Cakir}, provided that the fourth SM family exists. For same reasons, the SM
Higgs boson could be seen at the LHC via the ''golden mode'' even with an
integral luminosity of only a few fb$^{-1}$\cite{Arik2}. Therefore, the
presumed fourth family quarks will be discovered at LHC well before getting
CLIC into operation. The same statement is valid for pseudoscalar quarkonia (%
$\eta _{4}$) formed by the fourth family quarks \cite{ATLAS,ATLAS2}.
Observation of the fourth family leptons at LHC is problematic due to large
backgrounds from the production of real W and Z bosons, both singly and in
pairs, that submerge these heavy lepton signals \cite{ATLAS,Barger}. For
these reasons, lepton colliders will be advantageous for investigation of
the fourth SM family leptons and vector ($\psi _{4}$) quarkonia. The
potential of muon colliders in this context was analyzed in \cite{Çiftçi}.
With this paper we complete general overlook of\ the subject by considering
the potential of CLIC \cite{Assmann}.

\bigskip

\section{Flavor Democracy and the Standard Model}

\bigskip

It is usefull to consider three different bases:

- Standard Model basis $\left\{ f^{0}\right\} $,

- Mass basis $\left\{ f^{m}\right\} $,

- Weak basis $\left\{ f^{w}\right\} $.

According to the three family SM, before the spontaneous symmetry breaking
quarks are grouped into following SU(2)$\times $U(1) multiplets:

\begin{equation}
\left( 
\begin{array}{c}
u_{L}^{0} \\ 
d_{L}^{0}
\end{array}
\right) \text{, }u_{R}^{0}\text{, }d_{R}^{0}\text{; }\left( 
\begin{array}{c}
c_{L}^{0} \\ 
s_{L}^{0}
\end{array}
\right) \text{, }c_{R}^{0}\text{, }s_{R}^{0}\text{; }\left( 
\begin{array}{c}
t_{L}^{0} \\ 
b_{L}^{0}
\end{array}
\right) \text{, }t_{R}^{0}\text{, }b_{R}^{0}\text{ }
\end{equation}

In one family case all bases are equal and, for example, d-quark mass is
obtained due to Yukawa interaction

\begin{equation}
L_{Y}^{(d)}=a_{d}\left( 
\begin{array}{cc}
\overline{u}_{L} & \overline{d}_{L}
\end{array}
\right) \left( 
\begin{array}{c}
\varphi ^{+} \\ 
\varphi ^{-}
\end{array}
\right) d_{R}+h.c.\Rightarrow L_{m}^{(d)}=m_{d}\overline{d}d
\end{equation}
where $m_{d}=a_{d}\eta $, $\eta =\left\langle \varphi ^{0}\right\rangle
\cong 249$ GeV. In the same manner $m_{u}=a_{u}\eta $, $m_{e}=a_{e}\eta $
and $m_{\nu _{e}}=a_{\nu _{e}}\eta $ (if neutrino is Dirac particle). In $n$
family case

\begin{equation}
L_{Y}^{(d)}=\sum\limits_{i,j=1}^{n}a_{ij}^{d}\left( 
\begin{array}{cc}
\overline{u}_{Li}^{0} & \overline{d}_{Li}^{0}
\end{array}
\right) \left( 
\begin{array}{c}
\varphi ^{+} \\ 
\varphi ^{-}
\end{array}
\right) d_{Rj}^{0}+h.c.=\sum\limits_{i,j=1}^{n}m_{ij}^{d}\overline{d}%
_{i}^{0}d_{j}^{0}\text{, }m_{ij}^{d}=a_{ij}^{d}\eta
\end{equation}
where $d_{1}^{0}$ denotes $d^{0}$, $d_{2}^{0}$ denotes $s^{0}$ etc. The
diagonalization of mass matrix of each type of fermions, or in other words
transiton from SM basis to mass basis, is performed by well-known bi-unitary
transformation. Then, the transition from mass basis to weak basis result in
CKM matrix

\begin{equation}
U_{CKM}=\left( U_{L}^{u}\right) ^{+}U_{L}^{d}
\end{equation}
which contains 3(6) observable mixing angels and 1(3) observable
CP-violating phases in the case of three(four) SM families.

Before the spontaneous symmetry breaking all quarks are massless and there
are no differences between $d^{0}$, $s^{0}$ and $b^{0}$, fermions with the
same quantum numbers are indistinguishable. This leads us to the {\it first
assumtion }\cite{Harari}, namely, Yukawa couplings are equal within each
type of fermions:

\begin{equation}
a_{ij}^{d}\cong a^{d}\text{, }a_{ij}^{u}\cong a^{u}\text{, }a_{ij}^{l}\cong
a^{l}\text{, }a_{ij}^{\nu }\cong a^{\nu }
\end{equation}

The first assumtion result in $n-1$ massless particles and one massive
particle with $m=na^{F}\eta $ $\left( F=u\text{, }d\text{, }l\text{, }\nu
\right) $ for each type of the SM fermions.

Because there is only one Higgs doublet which gives Dirac masses to all four
types of fermions (up quarks, down quarks, charged leptons and neutrinos),
it seems natural to make the {\it second assumption }\cite{Datta,Celikel},
namely, Yukawa constants for different types of fermions should be nearly
equal:

\begin{equation}
a^{d}\approx a^{u}\approx a^{l}\approx a^{\nu }\approx a
\end{equation}

Taking into account the mass values for the third generation, the second
assumption leads to the statement that {\it according to the flavor
democracy the fourth SM family should exist}.

In terms of the mass matrix above arguments mean

\begin{equation}
M^{0}=a\eta \left( 
\begin{array}{cccc}
1 & 1 & 1 & 1 \\ 
1 & 1 & 1 & 1 \\ 
1 & 1 & 1 & 1 \\ 
1 & 1 & 1 & 1
\end{array}
\right) \Rightarrow M^{m}=4a\eta \left( 
\begin{array}{cccc}
0 & 0 & 0 & 0 \\ 
0 & 0 & 0 & 0 \\ 
0 & 0 & 0 & 0 \\ 
0 & 0 & 0 & 1
\end{array}
\right)
\end{equation}

Now, let us make the {\it third assumption}, namely, $a$ is between $%
e=g_{w}\sin \theta _{W}$ and $g_{w}/\cos \theta _{W}$. Therefore, the fourth
family fermions are almost degenerate, in good agreement with experimental
value $\rho =0.9998\pm 0.0008$ \cite{Erler}, and their common mass lies
between $320$ GeV and $730$ GeV. The last value is clouse to upper limit on
heavy quark masses, $m_{Q}\leq 700$ GeV, which follows from partial-wave
unitarity at high energies \cite{Chanowitz}. It is interest that with
preferable value $a\approx g_{w}$ flavor democracy predicts $m_{4}\approx
8m_{W}\approx 640$ GeV.

The masses of the first three family fermions, as well as an observable
interfamily mixings, are generated due to the small deviations from the full
flavor democracy \cite{Datta,Atag}. \ 

\bigskip

\section{\protect\bigskip Pair Production}

As a result of the DMM approach, the SM is extended to include a fourth
generation of fundamental fermions, with masses typically in the range from
300 GeV to 700 GeV. Therefore, CLIC with 1-3 TeV center of mass energy will
give opportunity to search the alleged fourth SM family fermions and
quarkonia in details.

\subsection{$e^{+}e^{-}$ option}

The annihilation of $e^{+}e^{-}$ is a classic channel to produce and study
new heavy fermions, because the cross sections are relatively large compared
with backgrounds \cite{Barger}. The cross section for the process $%
e^{+}e^{-}\rightarrow f$ $\stackrel{\_}{f}$ \ has the well-known form

\begin{equation}
\sigma =\frac{2\pi \alpha ^{2}}{3s}\xi \beta \left\{ Q_{f}\left( Q_{f}-2\chi
_{1}vv_{f}\right) \left( 3-\beta ^{2}\right) +\chi _{2}\left( 1+v^{2}\right) %
\left[ v_{f}^{2}\left( 3-\beta ^{2}\right) +2\beta ^{2}a_{f}^{2}\right]
\right\}
\end{equation}
where

$\chi _{1}=\frac{1}{16\sin ^{2}\theta _{W}\cos ^{2}\theta _{W}}\frac{s\left(
s-M_{Z}^{2}\right) }{\left( s-M_{Z}^{2}\right) ^{2}+\Gamma _{Z}^{2}M_{Z}^{2}}
$

$\chi _{2}=\frac{1}{256\sin ^{4}\theta _{W}\cos ^{4}\theta _{W}}\frac{s^{2}}{%
\left( s-M_{Z}^{2}\right) ^{2}+\Gamma _{Z}^{2}M_{Z}^{2}}$

$v=-1+4\sin ^{2}\theta _{W}$

$a_{f}=2T_{3f}$

$v_{f}=2T_{3f}-4Q_{f}\sin ^{2}\theta _{W}$

$\beta =\sqrt{1-4m_{Q}^{2}/s}$.

$T_{3}=\frac{1}{2}$ for $\nu _{4}$ and $u_{4}$, $T_{3}=-\frac{1}{2}$ for $%
l_{4}$ and $d_{4}$

$\xi =1$ for leptons, $\xi =3$ for quarks.

Obtained cross section values for pair production of the fourth SM family
fermions with $m_{4}=320$ ($640)$ GeV and corresponding number of events per
working year ($10^{7}$ s) are given in Table I (Table II). Event signatures
are defined by the mass pattern of the fourth family and $4\times 4$
Cabibbo-Kobayashi-Maskawa (CKM) matrix elements. According to scenario given
in \cite{Atag}, dominant decay modes are $u_{4}\rightarrow b$ $W^{-}$, $%
d_{4}\rightarrow t$ $W^{+}$, $l_{4}\rightarrow \nu _{\tau }W^{-}$ and $\nu
_{4}\rightarrow \tau ^{-}W^{+}$.

We have mentioned in the introduction that the fourth family quarks with $%
m_{4}<1$ TeV will be discovered at LHC. However, CLIC\ will raise an
opportunity to investigate their properties in details due to sufficiently
large event numbers and clean environment. The observation of the fourth
family leptons at hadron colliders is not very promising because of low
statistics and large background. The charged $l_{4}$ lepton will have clear
signature at CLIC. For example, if produced $W^{\pm }$ bosons decay
leptonically, one deals with two acoplanary opposite charge leptons and
large missing energy. Pair production of the neutral $\nu _{4}$ leptons will
lead to more complicated event topology. In this case, $\tau $ tagging will
be helpfull in identification of events. Indeed, produced $\tau $ leptons
will decay at the distance 1-2 cm from interaction point, which can be
easily measured by vertex detector.

In addition, polarization of electron and positron beams should help to
experimentally determine axial and vector neutral current coupling constants
of the fourth family fermions. Note that hadron colliders can not provide
sufficient information about them. This subject is under study.

\subsection{$\protect\gamma \protect\gamma $ option}

It is well known that linear $e^{+}e^{-}$\ colliders \ will give opportunity
to construct TeV energy $\gamma \gamma $ colliders on their basis \cite
{Ginzburg1,Ginzburg2,Telnov}. The fourth SM family quarks and charged
leptons will be copiously produced at $\gamma \gamma $ machines. The cross
section for $\gamma \gamma \rightarrow f\overline{f}$ at fixed $\widehat{s}$
has the form

\begin{equation}
\widehat{\sigma }=\frac{2\xi \pi \alpha _{em}^{2}Q_{f}^{4}}{\widehat{s}%
(1+\beta ^{2})}\left[ 2\beta (\beta ^{4}-\beta ^{2}-2)+(\beta ^{6}+\beta
^{4}-3\beta ^{2}-3)\ln \left( \frac{1-\beta }{1+\beta }\right) \right]
\end{equation}
where $\beta =\sqrt{1-4m^{2}/\widehat{s}}$. Since Compton backscattered
photons are not monochromatic, one should perform further integrations over
the photon spectrum to obtain the outcame cross section

\begin{equation}
\sigma =\int\limits_{\tau _{\min }}^{(0.83)^{2}}d\tau \int\limits_{\tau
/0.83}^{0.83}\frac{dx}{x}f_{\gamma }\left( \frac{\tau }{x}\right) f_{\gamma
}\left( x\right) \widehat{\sigma }\left( \tau s\right)
\end{equation}
where $\tau _{\min }=4m^{2}/s$ and $\widehat{s}=\tau s$. The energy spectrum
of the high energy photons obtained through Compton backscattering of laser
photons on the high energy electron beam has the form

\begin{equation}
f_{\gamma }(y)=\frac{1}{1.84}\left[ 1-y+\frac{1}{1-y}-\frac{4y}{\zeta (1-y)}+%
\frac{4y^{2}}{\zeta ^{2}(1-y)^{2}}\right]
\end{equation}
with $\zeta =4.8$. For $\sqrt{s_{ee}}=1$ TeV ($3$ TeV), which corresponds to 
$\sqrt{s_{\gamma \gamma }}^{\max }=0.83\sqrt{s_{ee}}=0.83$ TeV ($2.5$ TeV),
the obtained values of cross section and event number per year are presented
in Tables I (Table II). \bigskip

\section{Quarkonium Production}

\bigskip The condition for forming $(Q\overline{Q})$ quarkonia states with
new heavy quarks is \cite{Bigi}

\begin{equation}
m_{Q}\leq (125\text{ }GeV)\text{ }\left| V_{Qq}\right| ^{-2/3}
\end{equation}
where $q=d,s,b$ for $Q=u_{4}$ and $q=u,c,t$ for $Q=d_{4}$. Differing from t
quark, fourth family quarks will form quarkonia because $u_{4}$ and $d_{4}$
are almost degenerate and\ their decays are suppressed by small CKM mixings 
\cite{Celikel,Atag,Sultansoy}. Below, we consider resonance productions of $%
\psi _{4}$ quarkonia at $e^{+}e^{-}$ and $\eta _{4}$ quarkonia at $\gamma
\gamma $ options of CLIC.

\subsection{$e^{+}e^{-}$ option}

The cross section for the formation of the fourth family quarkonium is given
with the well-known relativistic Breit-Wigner equation 
\begin{equation}
\sigma \left( e^{+}e^{-}\rightarrow \left( Q\stackrel{\_}{Q}\right) \right) =%
\frac{12\pi \left( s/M^{2}\right) \Gamma _{ee}\Gamma }{\left( s-M^{2}\right)
^{2}+M^{2}\Gamma ^{2}},  \label{bir}
\end{equation}
where $M$ is the mass, $\Gamma _{ee}$ is the partial decay width to $%
e^{+}e^{-}$ and $\Gamma $ is the total decay width of the fourth family
quarkonium. Using corresponding formulas from \cite{Barger87} in the
framework of Coulomb potential model, we obtain decay widths for the main
decay modes of $\psi _{4}(u_{4}\overline{u_{4}})$ and $\psi _{4}(d_{4}%
\overline{d_{4}})$, which are given in Table III. One can see that dominant
decay mode for both $\psi _{4}(u_{4}\overline{u_{4}})$ and $\psi _{4}(d_{4}%
\overline{d_{4}})$ quarkonia is $\psi _{4}\rightarrow W^{+}W^{-}$. Next
important decay modes are $\psi _{4}\rightarrow \gamma Z$ and $\psi
_{4}\rightarrow \gamma H$.

In order to estimate the number of produced quarkonium states, one should
take into account the luminosity distrubition at CLIC which is influenced
from energy spread of electron and positron beams and beamstrahlung. In our
calculations we use GUINEA-PIG simulation code \cite{Schulte}. For
illustration, we assume that $m_{\psi _{4}}\simeq 1$ TeV. The estimated\
event numbers per year for $\psi _{4}$ production, as well as, $\psi
_{4}\rightarrow \gamma H$ and $\psi _{4}\rightarrow ZH$ decay channels are
presented in Table IV. In our opinion, $\gamma H$ decay mode of $\psi
_{4}(u_{4}\overline{u_{4}})$ quarkonium is promising for investigation of
Higgs boson properties, especially if energy spread of electron and positron
beams about $10^{-3}$ \ could be managed succesfully.

Since quarkonia are produced at resonance ($\sqrt{s}\simeq m_{\psi _{4}}$)
and the mass of the fourth family quarks are close to $m_{\psi _{4}}/2$, the
uncertainity in measurement of quarks' masses is determined by energy spread
of the colliding beams. Therefore, if $m_{4}=500$ GeV, $\Delta m=5$ GeV for $%
\Delta E/E=10^{-2}$ and $\Delta m=0.5$ GeV for $\Delta E/E=10^{-3}$. For
comparision at LHC $\Delta m=22$ GeV for $m_{4}=320$ GeV and $\Delta m=36$
GeV for $m_{4}=640$ GeV \cite{Arik,ATLAS}.

\subsection{$\protect\gamma \protect\gamma $ option}

Pseudoscalar $\eta _{4}$ quarkonia formed by the fourth SM family quarks
will be copiously produced at LHC \cite{ATLAS,ATLAS2}. Decay widths for main
decay modes of $\eta _{4}(u_{4}\overline{u_{4}})$ and $\eta _{4}(d_{4}%
\overline{d_{4}})$ are given in Table V. As seen from the table, the
dominant decay mode is $\eta _{4}\rightarrow ZH$. One can estimate $\gamma
\gamma \rightarrow \eta _{4}$ production cross section approximately by
using following relation \cite{Gorbunov}:

\begin{equation}
\sigma \approx 50\text{fb}(1+\lambda _{1}\lambda _{2})\left( \frac{\text{Br}%
_{\gamma \gamma }}{4\times 10^{-3}}\right) \left( \frac{\Gamma _{tot}}{1%
\text{ MeV}}\right) \left( \frac{200\text{ GeV}}{M_{\eta }}\right) ^{3}
\end{equation}
where Br$_{\gamma \gamma }$ is the branching ratio of $\eta _{4}\rightarrow
\gamma \gamma $ decay mode, $\Gamma _{tot}$ is the\ quarkonium total decay
width and $\lambda _{1,2}$ are helicities of initial photons. Assuming $%
\lambda _{1}\lambda _{2}=1$, we obtain total event numbers of $\eta _{4}$
production, as well as numbers of $\eta _{4}\rightarrow ZH$ events, which
are given in Table VI. The advantage of $\eta _{4}(u_{4}\overline{u_{4}})$
with respect to $\eta _{4}(d_{4}\overline{d_{4}})$ is obvious.\ \ \ 

\section{Conclusion}

\bigskip We have shown that both fourth family fermions and quarkonia will
be copiously produced at CLIC. If the fourth SM family exists, CLIC will be
excellent place for investigation of fourth family quarkonia and leptons.
Formation of the fourth SM family quarkonia will give a new opportunity to
investigate Higgs bosons properties especially due to $e^{+}e^{-}\rightarrow
\psi _{4}\rightarrow \gamma H$ channel. In addition, masses of $u_{4}$ and $%
d_{4}$ quarks will be measured with high accuracy.

It is possible that other lepton colliders such as NLC, TESLA or JLC can get
into operation before CLIC. Of course, the fourth family fermions can be
produced at these machines of kinematically allowed. The first stage of all
of them (including CLIC) is designed with $\sqrt{s}=0.5$ TeV. The second
stage is thought of $\sqrt{s}=1$ TeV with the exception of TESLA (with $%
\sqrt{s}=0.8$ TeV). Our result presented in Tables I, IV and VI are
applicable for second stage NLC and JLC with a suitable scaling factors
coming from luminosity and energy spread. Therefore, all linear
electron-positron colliders are almost equal as long as kinematics allows.
Finally, obvious\ advantage of CLIC is $3$ TeV option.

\section{Acknowledgements}

Authors are grateful to E. Ar\i k, O. \c{C}ak\i r, S. A. \c{C}etin, A. De
Roeck, D. Schulte, V. I. Telnov and \"{O}. Yava\c{s} for useful discussions.
This work is supported in part by Turkish Planning Organization (DPT) under
the Grant No 2002K120250.

\begin{table}[tbp] \centering%
%
\caption{Cross sections and event numbers per year for pair production of
the fourth standard model family fermions with mass 320 GeV at CLIC ($\sqrt{s_{ee}}=1$ TeV, $L_{ee}=2.7\times 10^{34}$cm$^{-2}$s$^{-1}$ and $L_{\gamma \gamma }=10^{34}$cm$^{-2}$s$^{-1}$)\label{key}}
\begin{tabular}{llllll}
&  & $u_{4}\overline{u_{4}}$ & $d_{4}\overline{d_{4}}$ & $l_{4}\overline{%
l_{4}}$ & $\nu _{4}\overline{\nu _{4}}$ \\ \hline
$e^{+}e^{-}$ option & \multicolumn{1}{|l}{$\sigma $ (fb)} & 130 & 60 & 86 & 
15 \\ 
& \multicolumn{1}{|l}{N$_{ev}$/year} & 35000 & 16000 & 23000 & 4100 \\ \hline
$\gamma \gamma $ option & \multicolumn{1}{|l}{$\sigma $ (fb)} & 34 & 2 & 58
& - \\ 
& \multicolumn{1}{|l}{N$_{ev}$/year} & 3400 & 200 & 5700 & -
\end{tabular}
\end{table}%
%

\begin{table}[tbp] \centering%
%
\caption{Cross sections and event numbers per year for pair production of
the fourth standard model family fermions with mass 640 GeV at CLIC ($\sqrt{s_{ee}}=3$ TeV, $L_{ee}=1\times 10^{35}$cm$^{-2}$s$^{-1}$ and $L_{\gamma \gamma }=3\times10^{34}$cm$^{-2}$s$^{-1}$)\label{key}}
\begin{tabular}{llllll}
&  & $u_{4}\overline{u_{4}}$ & $d_{4}\overline{d_{4}}$ & $l_{4}\overline{%
l_{4}}$ & $\nu _{4}\overline{\nu _{4}}$ \\ \hline
$e^{+}e^{-}$ option & \multicolumn{1}{|l}{$\sigma $ (fb)} & 16 & 8 & 10 & 2
\\ 
& \multicolumn{1}{|l}{N$_{ev}$/year} & 16000 & 8000 & 10000 & 2000 \\ \hline
$\gamma \gamma $ option & \multicolumn{1}{|l}{$\sigma $ (fb)} & 27 & 2 & 46
& - \\ 
& \multicolumn{1}{|l}{N$_{ev}$/year} & 8100 & 600 & 14000 & -
\end{tabular}
\end{table}%
%

\begin{table}[tbp] \centering%
%
\caption{Decay widths for main decay modes of $\psi _{4} $ for $m_{H}=150$
(300) GeV  with
$m_{\psi _{4}}\simeq 1$ TeV\label{key}} 
\begin{tabular}{lllll}
& \multicolumn{2}{c}{$\left( u_{4}\overline{u_{4}}\right) $} & 
\multicolumn{2}{c}{$\left( d_{4}\overline{d_{4}}\right) $} \\ \cline{2-5}
& $m_{H}=150$ GeV & $m_{H}=300$ GeV & $m_{H}=150$ GeV & $m_{H}=300$ GeV \\ 
\hline
$\Gamma (\psi _{4}\rightarrow \ell ^{+}\ell ^{-})$, 10$^{-3}$ MeV & 
\multicolumn{1}{c}{18.9} & \multicolumn{1}{c}{18.9} & \multicolumn{1}{c}{7.3}
& \multicolumn{1}{c}{7.3} \\ 
$\Gamma (\psi _{4}\rightarrow u\stackrel{\_}{u})$, 10$^{-2}$ MeV & 
\multicolumn{1}{c}{3.2} & \multicolumn{1}{c}{3.2} & \multicolumn{1}{c}{1.9}
& \multicolumn{1}{c}{1.9} \\ 
$\Gamma (\psi _{4}\rightarrow d\stackrel{\_}{d})$, 10$^{-2}$ MeV & 
\multicolumn{1}{c}{1.4} & \multicolumn{1}{c}{1.4} & \multicolumn{1}{c}{1.7}
& \multicolumn{1}{c}{1.7} \\ 
$\Gamma (\psi _{4}\rightarrow Z\gamma )$, 10$^{-1}$\ MeV & 
\multicolumn{1}{c}{15} & \multicolumn{1}{c}{15} & \multicolumn{1}{c}{3.7} & 
\multicolumn{1}{c}{3.7} \\ 
$\Gamma (\psi _{4}\rightarrow ZZ)$, 10$^{-1}$ MeV & \multicolumn{1}{c}{1.7}
& \multicolumn{1}{c}{1.7} & \multicolumn{1}{c}{5.4} & \multicolumn{1}{c}{5.4}
\\ 
$\Gamma (\psi _{4}\rightarrow ZH)$, 10$^{-1}$ MeV & \multicolumn{1}{c}{1.7}
& \multicolumn{1}{c}{1.6} & \multicolumn{1}{c}{5.5} & \multicolumn{1}{c}{5.2}
\\ 
$\Gamma (\psi _{4}\rightarrow \gamma H)$, 10$^{-1}$ MeV & \multicolumn{1}{c}{
14.4} & \multicolumn{1}{c}{13.4} & \multicolumn{1}{c}{3.6} & 
\multicolumn{1}{c}{3.4} \\ 
$\Gamma (\psi _{4}\rightarrow W^{+}W^{-})$, MeV & \multicolumn{1}{c}{70.8} & 
\multicolumn{1}{c}{70.8} & \multicolumn{1}{c}{71.2} & \multicolumn{1}{c}{71.2
}
\end{tabular}
\end{table}%
%

\begin{table}[tbp] \centering%
%
\caption{The production event numbers per year for the fourth SM family
$\psi_{4}$ quarkonia  at CLIC 1 TeV option with
$m_{\psi _{4}}\simeq 1 $ TeV\label{key}} 
\begin{tabular}{lllcc}
&  &  & \multicolumn{2}{c}{Events per year} \\ \cline{4-5}
&  & $\Delta E/E$ & $m_{H}=150$ GeV & $m_{H}=300$ GeV \\ \hline
$e^{+}e^{-}\rightarrow \psi _{4}$ & \multicolumn{1}{|l}{$\left( u_{4}%
\overline{u_{4}}\right) $} & $10^{-2}$ & 3500 & 3500 \\ 
& \multicolumn{1}{|l}{} & $10^{-3}$ & 26600 & 26700 \\ \cline{2-5}
& \multicolumn{1}{|l}{$\left( d_{4}\overline{d_{4}}\right) $} & $10^{-2}$ & 
1400 & 1400 \\ 
& \multicolumn{1}{|l}{} & $10^{-3}$ & 10400 & 10400 \\ \hline
$e^{+}e^{-}\rightarrow \psi _{4}\rightarrow \gamma H$ & \multicolumn{1}{|l}{$%
\left( u_{4}\overline{u_{4}}\right) $} & $10^{-2}$ & 70 & 60 \\ 
& \multicolumn{1}{|l}{} & $10^{-3}$ & 510 & 480 \\ \cline{2-5}
& \multicolumn{1}{|l}{$\left( d_{4}\overline{d_{4}}\right) $} & $10^{-2}$ & 7
& 6 \\ 
& \multicolumn{1}{|l}{} & $10^{-3}$ & 50 & 70 \\ \hline
$e^{+}e^{-}\rightarrow \psi _{4}\rightarrow ZH$ & \multicolumn{1}{|l}{$%
\left( u_{4}\overline{u_{4}}\right) $} & $10^{-2}$ & 8 & 7 \\ 
& \multicolumn{1}{|l}{} & $10^{-3}$ & 60 & 60 \\ \cline{2-5}
& \multicolumn{1}{|l}{$\left( d_{4}\overline{d_{4}}\right) $} & $10^{-2}$ & 
10 & 10 \\ 
& \multicolumn{1}{|l}{} & $10^{-3}$ & 80 & 70
\end{tabular}
\end{table}%
%

\begin{table}[tbp] \centering%
%
\caption{Decay widths for main decay modes of $\eta  _{4} $ for $m_{H}=150$
(300) GeV with
$m_{\eta _{4}}= 0.75 $ TeV\label{key}} 
\begin{tabular}{lllll}
& \multicolumn{2}{c}{$\left( u_{4}\overline{u_{4}}\right) $} & 
\multicolumn{2}{c}{$\left( d_{4}\overline{d_{4}}\right) $} \\ \cline{2-5}
& $m_{H}=150$ GeV & $m_{H}=300$ GeV & $m_{H}=150$ GeV & $m_{H}=300$ GeV \\ 
\hline
$\Gamma (\eta _{4}\rightarrow \gamma \gamma )$, 10$^{-3}$ MeV & 
\multicolumn{1}{c}{19.5} & \multicolumn{1}{c}{19.5} & \multicolumn{1}{c}{1.06
} & \multicolumn{1}{c}{1.06} \\ 
$\Gamma (\eta _{4}\rightarrow Z\gamma )$, 10$^{-3}$\ MeV & 
\multicolumn{1}{c}{4.6} & \multicolumn{1}{c}{4.6} & \multicolumn{1}{c}{3.7}
& \multicolumn{1}{c}{3.7} \\ 
$\Gamma (\eta _{4}\rightarrow ZZ)$, 10$^{-1}$\ MeV & \multicolumn{1}{c}{2.2}
& \multicolumn{1}{c}{2.2} & \multicolumn{1}{c}{2.8} & \multicolumn{1}{c}{2.8}
\\ 
$\Gamma (\eta _{4}\rightarrow gg)$,\ MeV & \multicolumn{1}{c}{5.1} & 
\multicolumn{1}{c}{5.1} & \multicolumn{1}{c}{5.1} & \multicolumn{1}{c}{5.1}
\\ 
$\Gamma (\eta _{4}\rightarrow ZH)$, MeV & \multicolumn{1}{c}{47.3} & 
\multicolumn{1}{c}{30.9} & \multicolumn{1}{c}{47.3} & \multicolumn{1}{c}{30.9
} \\ 
$\Gamma (\eta _{4}\rightarrow W^{+}W^{-})$, 10$^{-2}$ MeV & 
\multicolumn{1}{c}{5.7} & \multicolumn{1}{c}{5.7} & \multicolumn{1}{c}{5.7}
& \multicolumn{1}{c}{5.7} \\ 
$\Gamma (\eta _{4}\rightarrow t\overline{t})$,\ MeV & \multicolumn{1}{c}{16.4
} & \multicolumn{1}{c}{16.4} & \multicolumn{1}{c}{16.4} & \multicolumn{1}{c}{
16.4} \\ 
$\Gamma (\eta _{4}\rightarrow b\overline{b})$, 10$^{-2}$\ MeV & 
\multicolumn{1}{c}{1.0} & \multicolumn{1}{c}{1.0} & \multicolumn{1}{c}{1.0}
& \multicolumn{1}{c}{1.0}
\end{tabular}
\end{table}%
%

\begin{table}[tbp] \centering%
%
\caption{The production event numbers per year for the fourth SM family
$\eta_{4} $ quarkonia at $\gamma \gamma $ option with
$m_{\eta _{4}}= 0.75 $ TeV\label{key}} 
\begin{tabular}{llll}
&  & \multicolumn{2}{c}{Events per year} \\ \cline{3-4}
&  & $m_{H}=150$ GeV & $m_{H}=300$ GeV \\ \hline
$\gamma \gamma \rightarrow \eta _{4}$ & $\left( u_{4}\overline{u_{4}}\right) 
$ & \multicolumn{1}{c}{900} & \multicolumn{1}{c}{900} \\ 
& $\left( d_{4}\overline{d_{4}}\right) $ & \multicolumn{1}{c}{56} & 
\multicolumn{1}{c}{56} \\ \hline
$\gamma \gamma \rightarrow \eta _{4}\rightarrow ZH$ & $\left( u_{4}\overline{%
u_{4}}\right) $ & \multicolumn{1}{c}{610} & \multicolumn{1}{c}{520} \\ 
& $\left( d_{4}\overline{d_{4}}\right) $ & \multicolumn{1}{c}{38} & 
\multicolumn{1}{c}{33}
\end{tabular}
\end{table}%
%

\end{document}